\documentclass[twocolumn,10pt]{revtex4}%
\usepackage[paperwidth=210mm,paperheight=297mm,centering,hmargin=2cm,vmargin=2.5cm]{geometry}
\usepackage{amsfonts}
\usepackage{amsmath}
\usepackage{amssymb}
\usepackage{bm}
\usepackage{graphicx}
\usepackage{color}
\usepackage{soul}
\usepackage{epsfig}%
\usepackage[dvipsnames]{xcolor} 
  \definecolor{bleu_cite}{RGB}{0,0,255}
\usepackage[colorlinks=true,allcolors = black,citecolor=bleu_cite,  ]{hyperref} 
\usepackage{natbib}
\usepackage{siunitx}

\def\fd{\textcolor{black}}

\usepackage[normalem]{ulem} 

\begin{document}
\title{\fd{Dual density waves with neutral and charged dipolar excitons of GaAs bilayers}}

\author{Camille Lagoin$^1$, Stephan Suffit$^{2}$, Kirk  Baldwin$^3$, Loren Pfeiffer$^3$ and Fran\c{c}ois Dubin$^{1,4,\dag}$}
\affiliation{$^1$ Institut des Nanosciences de Paris, CNRS and Sorbonne Universit{\'e}, Paris, France}
\affiliation{$^2$ Laboratoire de Materiaux et Phenomenes Quantiques, Universite Paris Diderot, Paris, France}
\affiliation{$^3$ PRISM, Princeton Institute for the Science and Technology of Materials, Princeton University, Princeton, USA}
\affiliation{$^4$ Université Côte d’Azur -- CNRS, CRHEA, Valbonne, France}
\affiliation{$\dag$ francois$\_$dubin@icloud.com (corresponding author)}

\begin{abstract}
\textbf{\fd{Strongly correlated quantum particles in lattice potentials are the building blocks for a large variety of quantum insulators, for instance Mott phases and density waves breaking the lattice symmetry. Such collective states are accessible to bosonic and fermionic systems. To expand further the spectrum of accessible quantum matter phases, mixing both species is theoretically appealing, since density order then competes with phase separation. Here we manipulate such  Bose-Fermi mixture by confining neutral (boson-like) and charged (fermion-like) dipolar excitons in an artificial square lattice of a GaAs bilayer. At unitary lattice filling, strong inter- and intra-species interactions stabilise insulating phases when the fraction of charged excitons is around (1/3, 1/2, 2/3). We evidence that dual Bose-Fermi density waves are then realised, with species ordered in alternating stripes. Our observations highlight that dipolar excitons allow for controlled implementations of Bose-Fermi Hubbard models extended by off-site interactions.}}
\end{abstract}

\maketitle

\fd{Strongly interacting quantum particles confined in periodic potentials are specifically described by the Hubbard model (HM) \cite{Salomon_2010,Arovas_2022}. This Hamiltonian constitutes one of the most celebrated models of condensed matter physics}. For fermions, the HM has been extensively studied, underlying the current theoretical description of high temperature superconductivity \cite{Arovas_2022,Tranquada_2015}. For bosons, it has also raised considerable attention, triggered by pioneering experiments with ultra-cold atoms  \cite{Greiner_2002,Chin_2009}. 

Recently, the physics of the HM has been fostered by two-dimensional semiconductors. \fd{These offer new opportunities to study bosonic and fermionic realisations of Hubbard Hamiltonians. Triangular moiré lattices obtained by interfacing atomic monolayers provide a successful example for Fermi-Hubbard physics. Indeed, moiré lattices have enabled electronic Wigner crystals \cite{Li_21,Regan_20}, as well as a wide range of insulating density waves (DWs) at fractional lattice fillings \cite{Huang_21,Xu_20,Jin_21}. For bosons, the HM has been implemented with dipolar excitons that are Coulomb bound electrons and holes spatially separated in a GaAs bilayer \cite{Combescot_ROPP}. These are efficiently confined in gate-defined electrostatic lattices \cite{Butov_lattice,Lagoin_2020} and caracterised by their photoluminescence (PL) emission. Mott insulator and checkerboard solid were then reported at integer and half fillings respectively \cite{Lagoin_22,Lagoin_22b}, the latter phase evidencing the extended Bose-Hubbard model \cite{Baranov_2011,Dutta_2015}. }

\fd{Here, we implement the Bose-Fermi Hubbard Hamiltonian extended by off-site interactions, by confining dipolar excitons and holes in an electrostatic lattice of a GaAs bilayer. These two species experience strong attractions. Holes are thus captured in lattice sites occupied by excitons, yielding fermion-like charged excitons (CX). With a charge control down to an ultra-low residual doping level, we implement the Bose-Fermi HM by controlling neutral and charged excitons fillings, $\nu_X$ and $\nu_{CX}$ respectively. At $(\nu_X+\nu_{CX})=1$, we uncover incompressible states for fractional values of $\nu_{CX}$, around (1/3, 1/2, 2/3). We find spectroscopic evidence that neutral and charged excitons then implement alternating striped DWs. Thus, we confirm the rich variety of quantum insulators accessible to Bose-Fermi mixtures \cite{Buchler_2003,Maciej_2004,Hofstetter_2008,Sugawa_2011}.}


As illustrated in Fig.1.a, we study a field-effect device where \fd{an array of surface gate electrodes, polarised at $V_g$, applies a transverse electric field sinusoidally varying in the plane of a GaAs double quantum well. The latter confine electrons and holes, spatially separated in each layer and Coulomb bound to realise dipolar excitons. These are characterised by their permanent electric dipole, interacting with the applied field. Hence, we confine excitons in a 250 nm period square lattice, with around 250 $\mu$eV depth. Excitons then  have two accessible confined states (Wannier states -- WS) separated by 150 $\mu$eV (Ref.\cite{Lagoin_22b} and Methods). Moreover, note that excitons are optically injected, using a 100 ns long laser excitation repeated at 1 MHz. The laser excitation power $P$ controls the lattice filling and we study the photoluminescence (PL) radiated 300 ns after extinction of the laser pulse (Methods). This delay is around half the excitons effective lifetime. Hence, carriers are efficiently thermalised down to our lowest bath temperature (T=330 mK) \cite{Beian_2015}.  }

Figure 1.c reports the PL spectra radiated by our device, \fd{for different gate voltages and for a lattice filling around 1}. Let us first note that here and in the following the PL is averaged over a (3x2.5) $\mu m^2$ region, thereby including around 120 lattice sites (Methods). While at $V_g=-1.02$ V the top panel of Fig.1.c shows that the PL is mostly due to the recombination of dipolar excitons (\fd{high energy peak}), we observe that a second emission grows at lower energies when $V_g$ is reduced to about \fd{$-0.9935$} V (middle panel). By further decreasing $V_g$ to  $-0.96$ V, the bottom panel shows that the PL is mostly given by the low energy contribution.

Dual PL emissions have been previously reported for GaAs bilayers \cite{Dietl_2019,Bar_Joseph_21}. It was shown that the low energy line marks \fd{the strong attraction between excitons and excess carriers, yielding (composite) fermionic complexes usually referred to as charged excitons (CX). On the other hand,}  the higher energy \fd{PL line} signals the radiative recombination of neutral dipolar excitons. \fd{The energy separation between these two contributions, around 1 meV, well matches theoretical expectations \cite{Suris_2003,Witham_2018}. Moreover, we note that excess carriers can only be attributed to holes for our heterostructure characterised by ultra-high electronic mobilities \cite{Loren_21}}. Indeed, holes are trapped in the regions of the bilayer where a perpendicular field is applied, i.e. inside the lattice, whereas electrons are expelled out of this zone \cite{Kowalik_2012}. 

\fd{Neutral dipolar excitons are characterised by a linear dependence of their PL energy with $V_g$ \cite{Ronen_2007}, scaling like ($e.d|V_g|/L$). Here $e$ denotes the electron charge, $d$ the spatial separation between the center of the two quantum wells and $L$ the thickness of our field-effect device (Methods). In Fig.1.e we evidence this signature (blue), for $-1.02\lesssim V_g \lesssim-0.92 $ such that the lattice depth remains essentially unchanged (Methods). Furthermore, we note that the  PL energy of CX follows the same scaling (red). We then deduce that holes are captured in the lattice sites occupied by dipolar excitons. As a result  our device efficiently confines both neutral and charged dipolar excitons, in the same lattice potential. This conclusion is further confirmed by scrutinising the spectrum of the PL emitted by CX. Indeed, Fig.1.d underlines that for low lattice fillings the PL spectrum consists of two emission lines, separated by around 100 $\mu$eV. This magnitude agrees reasonably with the theoretically expected energy separation between the 2 WS accessible to an exciton complex made by two holes and one electron, and for around 250 $\mu$eV lattice depth (Methods).}

\fd{Figure 1.c and extended data Fig.1 evidence that the fraction of neutral and charged excitons are efficiently varied by the gate voltage. On the other hand, we verified that the overall lattice filling $(\nu_X+\nu_{CX})$ is mostly given by the  laser excitation power $P$ (extended data Fig.1). Experimentally, to set $\nu_X$ and $\nu_{CX}$ we first adjusted $V_g$ so that $\nu_{CX}\sim0$. The neutral excitons filling $\nu_X$ is then controlled by $P$, set for instance such that $\nu_X=1$ by realising a Mott insulator. For the chosen laser excitation power, $\nu_{CX}$ is then tuned by the gate voltage while $(\nu_X+\nu_{CX})$ remains mostly unchanged (extended data Fig.1). Note that neutral and charged exciton fillings are quantified by computing their PL integrated intensities (Methods). Thus, we possibly assess the purity of our device. Indeed, extended data Fig.2 shows that by suitably adjusting $V_g$ the PL due to CX can not be distinguished from the spectral noise for $\nu_X\approx0.5$. This shows that the concentration of \fd{residual} holes is bound to $4\cdot10^7$ cm$^{-2}$ (Methods). Remarkably, this value is comparable to the record residual doping measured for our heterostructure \cite{Loren_21}, a regime previously inaccessible to optical techniques.}

\fd{By confining tuneable concentrations of neutral and charged excitons, our lattice device can emulate Hubbard Hamiltonians, continuously from the bosonic to the fermionic regime. We have recently quantified the former one \cite{Lagoin_22,Lagoin_22b}. Below, we first implement the Fermi Hubbard model by equalising the fraction of excess holes and excitons. Combined inter-species attractions and fast carrier tunnelling (Methods) then ensure that the lattice is only filled with CX.}

\fd{Exploring the $(P,V_g)$ parameter space we found two specific combinations, namely (7.6 nW, -0.93 V) and (14 nW, -0.945 V), for which the CX compressibility $\kappa_{CX}$ is minimised, see white areas in Fig.2.a-b. In fact, varying either $V_g$ or $P$ from these coupled values leads to a two-fold increase of $\kappa_{CX}$, towards the level given by poissonian fluctuations, since excitons (blue area) or holes (red area) are then in excess. Furthermore, the two incompressible phases are characterised by a a lorentzian-like PL spectrum, given by our spectral resolution (Fig.2.b-d). Hence, we deduce that CX mostly occupy the same state in the lattice. In the charge neutral regime we verified that $\nu_X\approx1/2$ for P= 7.6 nW and 1 for P=14 nW. The spectra displayed in Fig.2.a-b then evidence two insulating states, at $\nu_{CX}\approx(1/2, 1)$, theoretically corresponding to MI and checkerboard solid respectively \cite{Arovas_2022}.}

\fd{As for neutral excitons \cite{Lagoin_22b}, from the difference $\Delta$ between the PL energies at unitary and half fillings we deduce the nearest neighbour (NN) interaction strength $V_{CX,CX}$  (Fig.1.b). Indeed, at $\nu_{CX}=1$ the energy of a MI is enhanced by 4$V_{CX,CX}$ compared to a checkerboard phase, since NN couplings are fully avoided for the latter. Taking into account the slight difference of gate voltages necessary to realise the two insulating states, we obtain $\Delta=(-0.015 \cdot e.d/L+4\cdot V_{CX,CX})$. This leads  to $V_{CX,CX}=$ (140 $\pm15$) $\mu$eV. Importantly, this magnitude is confirmed by studying the thermal melting of the insulating state at $\nu_{CX}=1/2$. Indeed, extended data Fig.3 evidences that $\kappa_{CX}$ is minimised up to a critical temperature around 1 K. Thus, our measurements follow theoretical expectations, namely that a checkerboard solid melts when $k_BT$ becomes of the order of $V_{CX,CX}$ \cite{Lagoin_22b,Troyer_2002}.}

\fd{In the following, we turn to the Bose-Fermi HM in the regime where the overall filling $(\nu_X+\nu_{CX})$ is set to unity, since strong on-site interactions prevent double occupancies in our lattice (Methods). We continuously vary neutral and charged excitons filling fractions by sweeping the gate voltage, and we monitor simultaneously the corresponding compressibility, $\kappa_X$ and $\kappa_{CX}$ respectively, by statistically computing PL intensity fluctuations (Methods).}   

Figure 3.a presents $\kappa_X$  (blue)  and $\kappa_{CX}$ (red), normalised by the level set by poissonian fluctuations, as a function of $\nu_{CX}$. For $\nu_{CX}\lesssim0.15$, the PL spectrum is dominated by the excitonic component and $\kappa_X$  is minimised. Excitons then realise a boson-like MI \fd{that is weakly perturbed by the fraction of fermionic defects, i.e. by the fraction of charged excitons $\nu_{CX}$. Similarly, $\kappa_{CX}$ is minimised for $\nu_{CX}\gtrsim0.85$  so that charged excitons realise a fermion-like MI at unitary filling, weakly perturbed by the fraction of (bosonic defects) excitons $\nu_X$. By contrast, $\kappa_X$ and $\kappa_{CX}$ are mostly of the order of the level given by poissonian fluctuations for $0.2\lesssim\nu_{CX}\lesssim0.8$. The system evolves then in a normal fluid phase. Nevertheless, for $\nu_{CX}=$(0.35, 0.52, 0.61) $\pm$ 0.03 we strikingly observe that  $\kappa_X$ and $\kappa_{CX}$ simultaneously drop, manifesting correlated incompressible states.} 

\fd{Combined insulating phases at $\nu_{CX}\sim$ (1/3, 1/2, 2/3) suggest that neutral and charged excitons realise dual DWs \cite{Buchler_2003,Maciej_2004,Hofstetter_2008}. To confirm this expectation, we studied $\Delta E_{X,CX}(\nu_{CX})=(E_X-E_{CX}$), $E_X$ and $E_{CX}$ being the PL energy of neutral and charged excitons respectively. $\Delta E_{X,CX}$ quantifies the average difference between the surroundings of each specie, since NN couplings are the main interaction channel. Hence, $\Delta E_{X,CX}$ distinguishes the two types of orders that compete in the insulating regime \cite{Buchler_2003,Hofstetter_2008}, namely DWs and spatially separated Mott phases of neutral and charged excitons (Methods).}

\fd{We first extract the inter-species NN interaction strength $V_{X,CX}$ (Fig.1.b) by comparing $\Delta E_{X,CX}$ between  the two MI regimes (see extended data Fig.4). Indeed, at $\nu_{CX}\sim$ 0.1 neutral excitons realise a Mott insulator with a few charged excitons impurities, so that $\Delta E_{X,CX}(0.1)\sim4(V_{X,X}-V_{X,CX})$. In the same way, at $\nu_{CX}\sim0.9$  we obtain $\Delta E_{X,CX}(0.9)\sim4(V_{X,CX}-V_{CX,CX})$. Accordingly we find $V_{X,CX}=(V_{X,X}+V_{CX,CX})/2-(\Delta E_{X,CX}(0.1)-\Delta E_{X,CX}(0.9))/8$. From the measurements shown in Fig.3.b (black points), we conclude that $V_{X,CX}=(75\pm10)$ $\mu$eV, since $V_{X,X}= (30\pm5)$ $\mu$eV \cite{Lagoin_22b} and $V_{CX,CX}= (140\pm15$) $\mu$eV.}

\fd{As shown above, repulsive NN interactions between charged excitons greatly exceed other interaction strengths. As a result, Fig.3b and extended data Fig.4 show that $\Delta E_{X,CX}$ (dashed line in Fig.3b) is smallest for spatially separated Mott phases, notably at $\nu_{CX}=$ (1/3, 1/2, 2/3). By contrast, the greatest amplitude is obtained for a double checkerboard density wave at $\nu_{CX}=$ 1/2, while intermediate values are found for striped DWs (Fig.3.b and extended data Fig.4).  For this latter ordering, one expects at $\nu_{CX}\approx 2/3$ that $\Delta E_{X,CX}\sim2(V_{X,CX}-V_{CX,CX})$, corresponding to an increase by $2(V_{CX,CX}-V_{X,CX})\approx130$ $\mu$eV compared to the exciton MI regime. Remarkably, this enhancement is confirmed experimentally in Fig.3b, since ($\Delta E_{X,CX}(0.61)-\Delta E_{X,CX}(0.9))=$ $(100\pm15)$ $\mu$eV. Hence, we deduce that alternating stripe phases are realised at $\nu_{CX}=0.61\pm0.03$, relatively close to the theoretical 2/3 filling. Similarly, we verify in Fig.3.b that ($\Delta E_{X,CX}$(0.36)-$\Delta E_{X,CX}(0.12))\sim$ $2(V_{X,CX}-V_{X,X})$. This matching again reveals that alternating stripes are favoured, with a pattern symmetric to that for $\nu_{CX}\sim2/3$. Finally, at $\nu_{CX}\sim1/2$ Fig.3.b shows that ($\Delta E_{X,CX}$(0.52)$-\Delta E_{X,CX}$(0.12))$=(0\pm15)$ $\mu$eV, as expected for horizontal or vertical stripes. Otherwise ($\Delta E_{X,CX}$(0.52)$-\Delta E_{X,CX}$(0.1)) would amount to around 220 $\mu$eV (dotted line) for alternating checkerboards (Fig.3.b and extended data Fig.4).}


\fd{To conclude, we have characterised the insulating part of the phase diagram for the Bose-Fermi Hubbard model. Dipolar excitons naturally explore the regime  extended by off-site interactions, so that  density waves emerge, even at half boson/fermion filling unlike for mixtures bound to short-range interactions \cite{Buchler_2003,Hofstetter_2008}. Reducing further the carriers temperature to a few tens of milli-Kelvin, which is within experimental reach, would extend our studies to the situation where excitons exhibit extended phase coherence. Mixtures of neutral and charged excitons could then serve as a platform to engineer supersolid-like states, stabilised by ordered fermionic phases. Also, the low (boson) doping regime relevant to superconducting electronic systems may be accessed, by doping Mott insulators of charged excitons with a coherent fluid of neutral ones.}

\section*{Acknowledgments}
Research at CNRS (C.L. and F.D.) has been financially supported by IXTASE from the French Agency for Research (ANR-20-CE30-0032-01). The work at Princeton University (L.P. and K.B.) was funded in part by the Gordon and Betty Moore Foundation’s EPiQS Initiative, Grant GBMF9615 to L. Pfeiffer, and by the National Science Foundation MRSEC grant DMR 2011750 to Princeton University.

\section*{Authors contributions}

K.B. and L.P. realised the GaAs bilayer while C.L., S.S. and F.D. fabricated the gate electrodes imprinting the 250 nm period electrostatic lattice. C.L. and F.D. performed all experiments, data analysis, and wrote the manuscript. F.D. designed the project.

\section*{Competing financial interests}

The authors declare no financial interests.

\newpage

\onecolumngrid

\clearpage

\pagebreak

\begin{figure}[!ht]
\includegraphics[width=\linewidth]{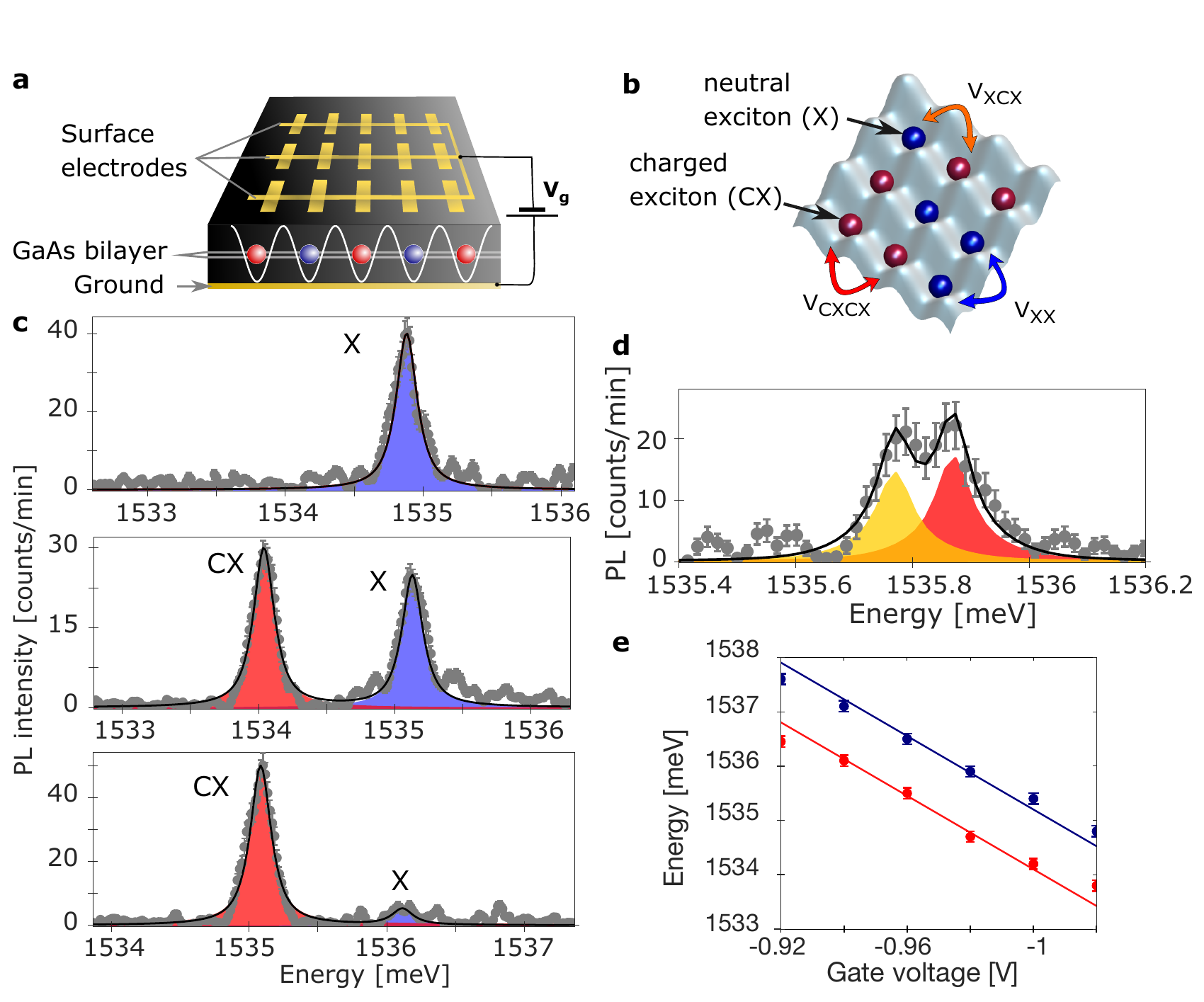}
  \caption{\textbf{\fd{Neutral and charged excitons Bose-Fermi mixture.}} \textbf{a} An array of gate electrodes polarised at $V_g$ imprints an artificial electrostatic 2D lattice (sinusoidal white wave) for neutral (blue) and charged (red) excitons confined in a GaAs bilayer (gray lines) embedded in an AlGaAs heterostructure (black). \textbf{b} In the lattice on-site interactions exceed the confinement depth so that neutral and charged excitons experience NN interactions \fd{only}, $V_{X,X}$, $V_{X,CX}$ and $V_{CX,CX}$. \textbf{c} PL spectra for \fd{$V_g=(-1.02, -0.9935, -0.96)$ V at unitary filling, from top to bottom. Lines represent model spectra obtained by adjusting the PL lines with a lorentzian-like profile limited by our spectral resolution, blue and red for neutral and charged excitons respectively.} \textbf{d} CX PL spectrum at half-filling reproduced by setting 44 and 56$\%$ occupations for the lower and upper WS, yellow and red respectively. The WS energy separation is set to 100 $\mu$eV while the black line displays the modelled profile. \textbf{e} Scaling of the charged and neutral exciton energies as a function of $V_g$, red and blue respectively. Points are wider than the horizontal error while the vertical error provides our spectral precision ($\pm15$ $\mu$eV). The solid lines display theoretical variations given by ($-e.d|V_g|/L$) (Methods). Experiments have all been realised at $T=330$ mK. In \textbf{c} and \textbf{d} error bars are given by the poissonian noise.}
  \label{fig:fig1}
\end{figure}

\newpage

\begin{figure}[!ht]
\includegraphics[width=\linewidth]{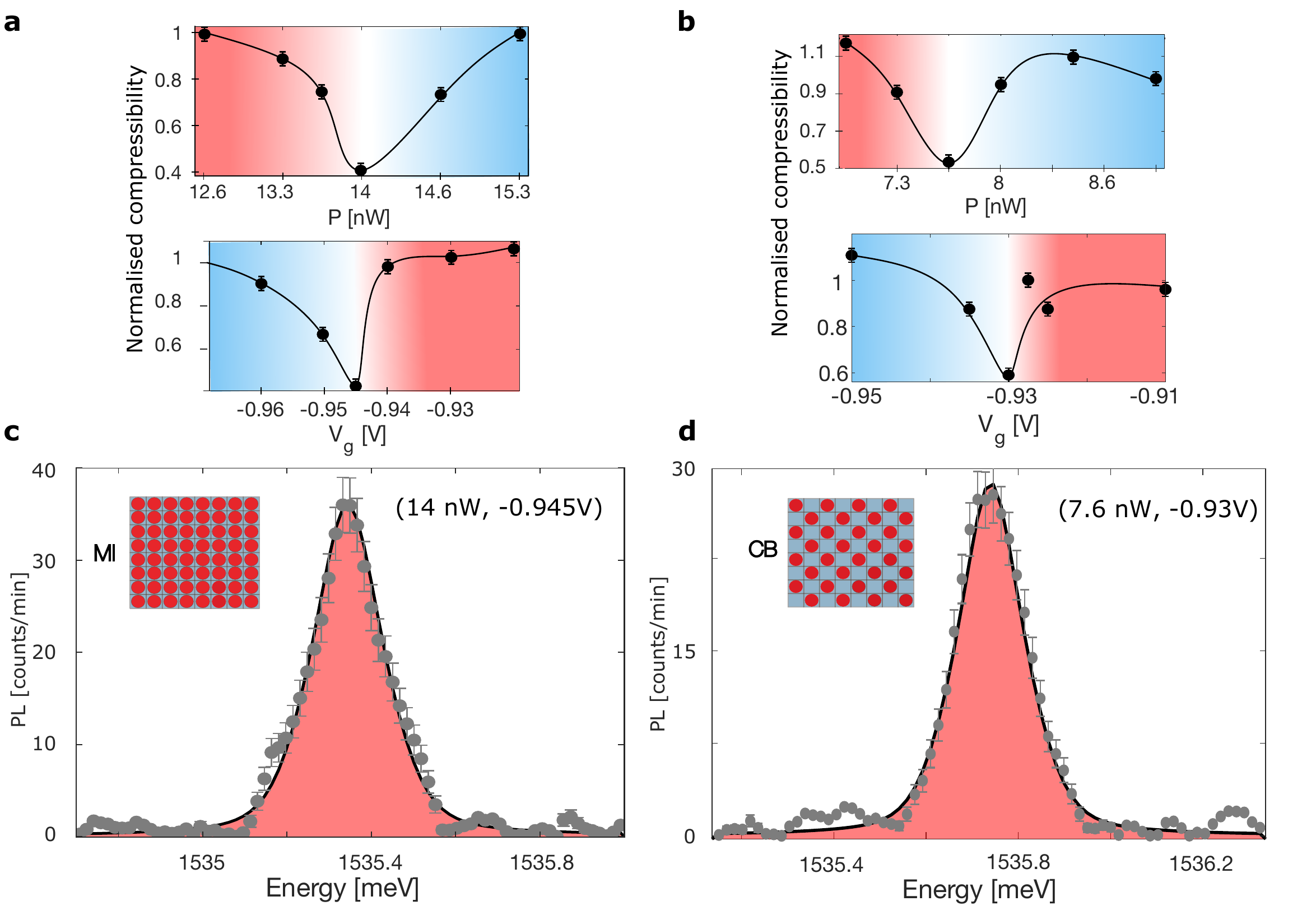}
  \caption{\textbf{\fd{Charged excitons insulators at unity and half fillings.}} \fd{\textbf{a}-\textbf{b} CX compressibility $\kappa_{CX}$ normalised to the level set by poissonian fluctuations, around unitary (\textbf{a}) and half filling (\textbf{b}). Bottom panels display the variation of $\kappa_{CX}$ as a function of $V_g$ for $P$=14 and 7.6 nW (\textbf{a} and \textbf{b} respectively). Top panels show $\kappa_{CX}$ as a function of $P$  for $V_g$=-0.945V (\textbf{a}) and -0.93V (\textbf{b}). The white area shows the regime where the hole and exciton densities are equalised so that the lattice only confines CXs. The blue and red areas show the regimes where excitons and holes are in excess, respectively. Black lines are guides for the eyes. \textbf{c}-\textbf{d} PL spectrum for $\nu_{CX}$ = 1 (14nW, -0.945V) and $\nu_{CX}$ = 1/2 (7.6nW, -0.93V), reproduced by the lorentzian-like profile given by our spectral resolution (red). Experiments have all been realised at $T=330$ mK, and we statistically analyse 10 repetitions for every experimental settings. In \textbf{c}-\textbf{d} error bars show the poissonian noise. In \textbf{a} and \textbf{b} vertical error bars provide the $\pm0.03$ precision when measuring the compressibility, while horizontal error bars are smaller than the points size.}}
  \label{fig:fig1}
\end{figure}

\newpage

\begin{figure}[!ht]
\includegraphics[width=\linewidth]{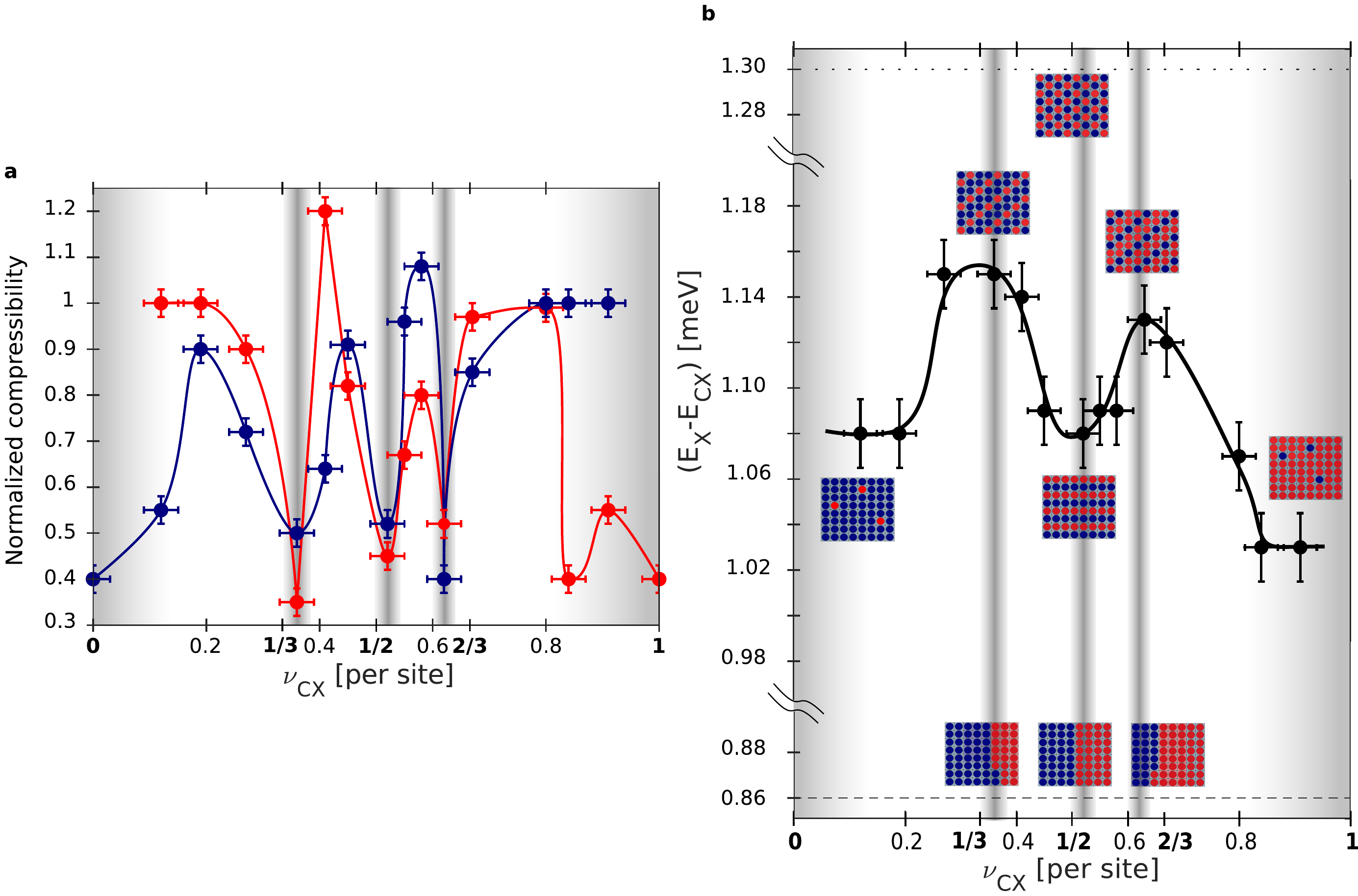}
  \caption{\textbf{\fd{Dual density waves of neutral and charged excitons.}} \fd{\textbf{a}} Charged (red) and neutral (blue) excitons compressibility as a function of \fd{$\nu_{CX}$ for $(\nu_X+\nu_{CX})=1$}. The compressibility is normalised to the value given by poissonian fluctuations. \fd{\textbf{b} Difference between the PL energy of neutral ($E_X$) and charged ($E_{CX}$) excitons as a function of $\nu_{CX}$. Black points represent our experimental data while the black line is a guide for the eyes. At the bottom, the horizontal dashed line shows that spatially separated insulators of neutral and charged excitons yield $(E_X-E_{CX})$=0.86 meV at $\nu_{CX}$=(1/3,1/2,2/3), while a double checkerboard DW at $\nu_{CX}$=1/2 leads to $(E_X-E_{CX})$=1.30 meV (top horizontal dotted line).  Ordered phases are shown with neutral and charged excitons depicted by blue and red balls respectively.} In \textbf{a}-\textbf{b} grey areas underline insulating phases. Experiments have all been realised at $T=330$ mK, computing 10 repetitions for every experimental settings. \fd{Error bars signal our instrumental precision for PL energies ($\pm15$ $\mu$eV), $\nu_{CX}$ ($\pm0.03$) and the compressibility ($\pm0.03$).}  }
  \label{fig:fig1}
\end{figure}

\newpage

\newpage

\onecolumngrid

\newpage

\twocolumngrid

\section*{Methods}

\subsection{Sample structure and experimental procedure}

The \SI{250}{\nano\metre} period electrostatic lattice is identical to the one studied in Ref. \cite{Lagoin_22b}. It is based on two \SI{8}{\nano\metre} wide GaAs quantum wells, separated by a \SI{4}{\nano\metre} AlGaAs barrier. Dipolar excitons made by spatially separated electrons and holes then carry a permanent electric dipole with a moment $d$ around 12$e$.nm, where $e$ denotes the electron charge. Also, note that the quantum wells are positioned \SI{200}{\nano\metre} below the surface of the field-effect device where they are embedded, and \SI{150}{\nano\metre} above a conductive layer that serves as electrical ground. 

Electronic carriers are injected in the lattice potential using a laser excitation at resonance with the direct exciton absorption of the quantum wells. Hence, we ensure that the concentration of photo-injected free carriers is minimised, \fd{nevertheless not necessarily vanishing. This regime is only found by suitably adjusting $V_g$. In fact charge control is accessed for correlated values of $P$ and $V_g$. This behaviour is illustrated Fig.2, since the optimal values of $V_g$ to engineer CX insulators at $\nu_{CX}$= 1 and 1/2 differ by 0.015 V.}

To spectrally analyse the PL radiated by our device we rely on a 1800 lines/mm grating. The PL is then sampled with 15 $\mu$eV precision so that the narrowest profile possibly measured, e.g. a laser line, is lorentzian-like with around 150 $\mu$eV full-width-at-half-maximum. To model PL spectra we assign this profile to the emission from the WS for neutral and charged excitons.

\fd{As stated in the main text, we extract $\nu_{CX}$ and $\nu_{X}$ by computing the PL integrated intensities of neutral and charged excitons, these being directly proportional to the concentration of each species. Importantly, we verified that the overall lattice filling ($\nu_X+\nu_{CX}$) weakly varies in the explored range of gate voltages. Indeed, extended data Fig.1 shows that for $P$ fixed the overall PL integrated intensity is rather constant. Accordingly, we assumed that only $\nu_X$ and $\nu_{CX}$ vary with $V_g$, and given the signal-to-noise-ratio of our experiments we deduced that $\nu_{CX}$ is obtained with $\pm0.03$ precision. }

\subsection{\fd{Excitonic compressibility}}

\fd{To quantify the compressibility of neutral and charged excitons} we statistically analyse intensity fluctuations of their PL maxima, $A_{X}^{(\mathrm{max})}$ and $A_{CX}^{(\mathrm{max})}$ respectively. In practice, we rely on samples of 10 measurements all performed under fixed conditions, for every experimental settings. Note that the sample size is limited by the time (6h) available at our lowest bath temperature, since acquiring a single 10 measurements sample requires around 20 minutes. 

By monitoring PL intensity fluctuations, we directly compute the compressibility, since $\sqrt{\kappa_{X,CX}k_BT}$ is proportional to $\sigma(A_{X,CX}^{(\mathrm{max})})/\overline{A_{X,CX}^{(\mathrm{max})}}$ according to the fluctuation-dissipation theorem \cite{Chin_2009}. Here $\sigma(A_{X,CX}^{(\mathrm{max})})$ and $\overline{A_{X,CX}^{(\mathrm{max})}}$ denote the variance and the mean value of the PL maximum $A_{X,CX}^{(\mathrm{max})}$ respectively. Moreover, note that we systematically rescale neutral and charged excitons compressibility to the value computed for the level of poissonian fluctuations, for every experimental settings. \fd{Thus, experimental observations are more easily confronted.  Finally, for our sample size, i.e. 10 realisations for each experimental conditions, we estimated the precision with which the compressibility is deduced. For that, we computed $\kappa_{X,CX}$ by randomly selecting 7 measurements out of 10. Hence, we observed that $\kappa_{X,CX}$ varies by around $\pm0.03$. We then assigned this error bar to measured compressibility.}

\subsection{Neutral and charged excitons in the lattice}

Relying on finite element simulations we have computed the lattice profile \fd{confining neutral dipolar excitons}. We thus found that the lattice depth $V_0$ is about 250 $\mu$eV when $V_g$ is set to  $-1$V. Then, excitons are confined in two WS, separated by around 150 $\mu$eV. \fd{Furthermore, we computed the variation of the lattice depth when $V_g$ is varied between $-0.92$ and $-1.02$ V, as in Figs.1-3. We concluded that $V_0$ decreases from 250 to 230 $\mu$eV between these two values, and that the energy separation between the two accessible WS ranges then from 140 to 150 $\mu$eV. This variation is small compared to other energy scales, such as the thermal energy. Then it plays a minimal role in our studies.}  

\fd{We have also simulated the nature of the lattice confinement for CX. Noting that holes have their minimum potential energy at the position of the lattice sites, we first considered that holes are captured in lattice sites occupied by dipolar excitons, favoured by the strong inter-species attraction. For electron and hole effective masses, $m_e=0.07m_0$ and $m_h=0.12 m_0$ \cite{Witham_2018} respectively, and considering the same 250 $\mu$eV deep lattice, we find that CX are confined in two WS separated by around 120 $\mu$eV, in reasonable agreement with experimental findings (Fig.1).} 

\fd{Alternatively, one may consider that CX are confined in the lattice due to the interaction between their positive charge and the electric field applied in the plane of the double GaAs quantum wells. To explore this scenario} we computed the lattice depth for a point-like and positively charged exciton.  We deduced that $V_0$ would then be around 1 meV, but more importantly that the point-like charged particle would have access to 3 WS, separated by around 250 $\mu$eV. This conclusion is in contradiction with the measurements shown in Fig.1. The hole confinement in the lattice is then better captured by considering their strong Coulomb attraction with dipolar excitons.

\subsection{Hubbard parameters for neutral and charged excitons}
We have recently calculated and measured the on-site  and NN interactions between dipolar excitons, $U_{X,X}$ and $V_{X,X}$ respectively \cite{Lagoin_22,Lagoin_22b}. We thus found that $U_{XX}$ exceeds the lattice depth, it can be as large as 1 meV for the lowest WS, whereas $V_{X,X}=(30\pm5)$ $\mu$eV. \fd{On the other hand, the measurements shown in Fig.2-3 yield $V_{CX,CX}=(140\pm15)$ $\mu$eV and $V_{X,CX}=(75\pm10)$ $\mu$eV. This evidences that repulsions between neutral excitons constitute the weakest interaction channel. Also, we note that on-site intra- and inter-species scatterings largely exceed the lattice depth so that lattice sites can not be doubly occupied.}

\fd{In Ref.\cite{Lagoin_22b} we have shown that the tunnelling strength between lattice sites amounts to a few $\mu$eV for neutral excitons. Given that the hole effective mass is around 2/3 of the exciton one, we deduce that holes benefit from a tunnelling strength in the $\mu$eV range too. Thus we ensure that the lattice only confines CX when the hole and neutral excitons concentration are matched (Fig.2), since our experiments are acquired during a 100 ns long time interval allowing for hundreds of particle tunneling between lattice sites.}

\subsection{\fd{Minimum hole doping level}}

\fd{Extended data Fig.2 reports a PL spectrum measured for $\nu_{X}\approx1/2$ and for a minimised concentration of holes. The spectrum is obtained by averaging 10 realisations. Its maximum is ($34\pm1.8)$  counts/min while the level of spectral white noise is $(3.5\pm0.6)$ counts/min. The signal-to-noise-ratio ($SN$) is then about 10 in these experiments. Furthermore, we deduce that $\nu_{X}\approx1/2$ translates into a mean concentration of neutral excitons  $n_X\approx4\cdot10^8$ cm$^{-2}$ for our 250 nm period lattice. Since we do not distinguish the PL due to charged excitons from the spectral noise level, we deduce an upper limit for the mean concentration of holes in these measurements, namely $n_h\sim n_X/SN\approx4\cdot10^7$ cm$^{-2}$. This value, limited by $SN$, is then of the same order as the one obtained from transport techniques for our GaAs quantum wells ($\sim1.5\cdot10^7$ cm$^{-2}$, see Ref.\cite{Loren_21}).}

\subsection{\fd{Spatial order of incompressible phases}}

\fd{Within the framework of the Hubbard model, one expects theoretically that density waves emerge for fractional fillings, mostly around (1/3, 1/2, 2/3) where stripes or checkerboards orders are possibly realised. In our lattice neutral and charged excitons interact through NN couplings, $V_{XX}\sim V_{X,CX}/2\sim V_{CX,CX}/5$. We then directly deduce the mean single particle interaction energy, $\epsilon=(\nu_X\epsilon_X+\nu_{CX}\epsilon_{CX})$, $\epsilon_X$ and $\epsilon_{CX}$ denoting the mean neutral and charged excitons interaction energies respectively. }

\fd{At $\nu_{CX}\sim(1/3,2/3)$ DWs necessarily correspond to alternating diagonal stripes of neutral and charged excitons. We find that these are characterised by $\epsilon$= 4/3($2V_{X,CX}+V_{X,X}$)$\approx$ 240 $\mu$eV and 4/3($2V_{X,CX}+V_{CX,CX})$)$\approx$ 387 $\mu$eV at $\nu_{CX}=$ 1/3 and 2/3 respectively (see extended data Fig.4). On the other hand,  phase separated Mott insulators exhibit mean interaction energies around 267 and 413 $\mu$eV respectively. These are thereby unfavourable. The previous explicit evaluations illustrate that alternating stripe phases exhibit lower energies than separated Mott phases when $V_{X,CX}<(V_{X,X}+V_{CX,CX})/2$. Importantly this condition is satisfied in our experiments, since $V_{X,CX}=(V_{X,X}+V_{CX,CX})/2-(\Delta E_{X,CX}(0.1)-\Delta E_{X,CX}(0.9))/8$ as detailed in the main text.}

\fd{At $\nu_{CX}\sim1/2$,  we deduce that $\epsilon=4V_{X,CX}\approx300$ $\mu$eV for a double checkerboard DW, and $(2V_{X,CX}+V_{X,X}+V_{CX,CX})\approx 320$  $\mu$eV for a striped DW. On the other hand $\epsilon=2(V_{X,X}+V_{CX,CX})\approx340$ $\mu$eV for separated Mott phases of neutral and charged excitons. Again, we recover that DW arrangements are energetically favoured over phase separation. However, we also find that the alternating stripe phase exhibits a larger mean interaction energy than the checkerboard one. This conclusion contrasts with our observations. This may first suggest that the precision with which we extract NN interaction strengths is not sufficient to accurately determine the DW ground-state energy. Alternatively, one can not exclude that the alternating stripe phase at $\nu_{CX}$=1/2 is of metastable nature. Indeed, insulating states with Bose-Fermi mixtures are characterised by instabilities, triggered by a sensitive competition between inter- and intra-species interaction strengths \cite{Buchler_2003,Maciej_2004,Hofstetter_2008}.}

\newpage

\onecolumngrid

\centerline{\includegraphics[width=\linewidth]{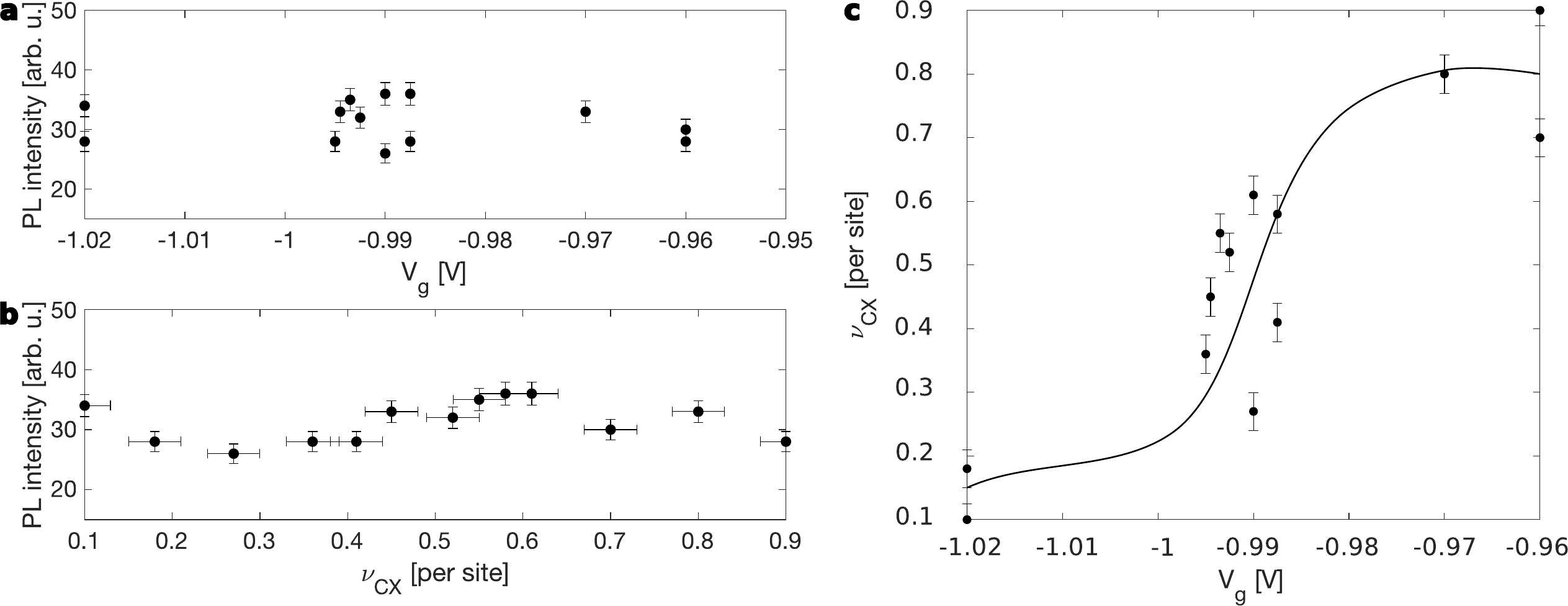}}
\textbf{\fd{Extended data Fig.1}}: \fd{\textbf{Lattice filling vs. gate voltage.}} 
\fd{\textbf{a} Total integrated intensities of the PL radiated by neutral and charged excitons, as a function of $V_g$ and at unitary filling ($P$=14 nW). \textbf{b} Same experimental results as in \textbf{a} but expressed as a function of $\nu_{CX}$.
\textbf{c} Scaling of $\nu_{CX}$ as a function of $V_g$ deduced from the measurements shown in \textbf{a} and \textbf{b}. The line provides a guide for the eyes. Experiments were all realised at 330 mK and acquired during four different experimental runs so that detection efficiencies are close but not identical. Vertical error bars display the poissonian precision in \textbf{a}-\textbf{b} and the $\pm 0.03$ precision on $\nu_{CX}$ in \textbf{c}. In \textbf{a}-\textbf{c}, the horizontal error is smaller than the points size while in \textbf{b} it corresponds to the precision when extracting $\nu_{CX}$.}

\newpage

\vspace{1cm}

\centerline{\includegraphics[width=.6\linewidth]{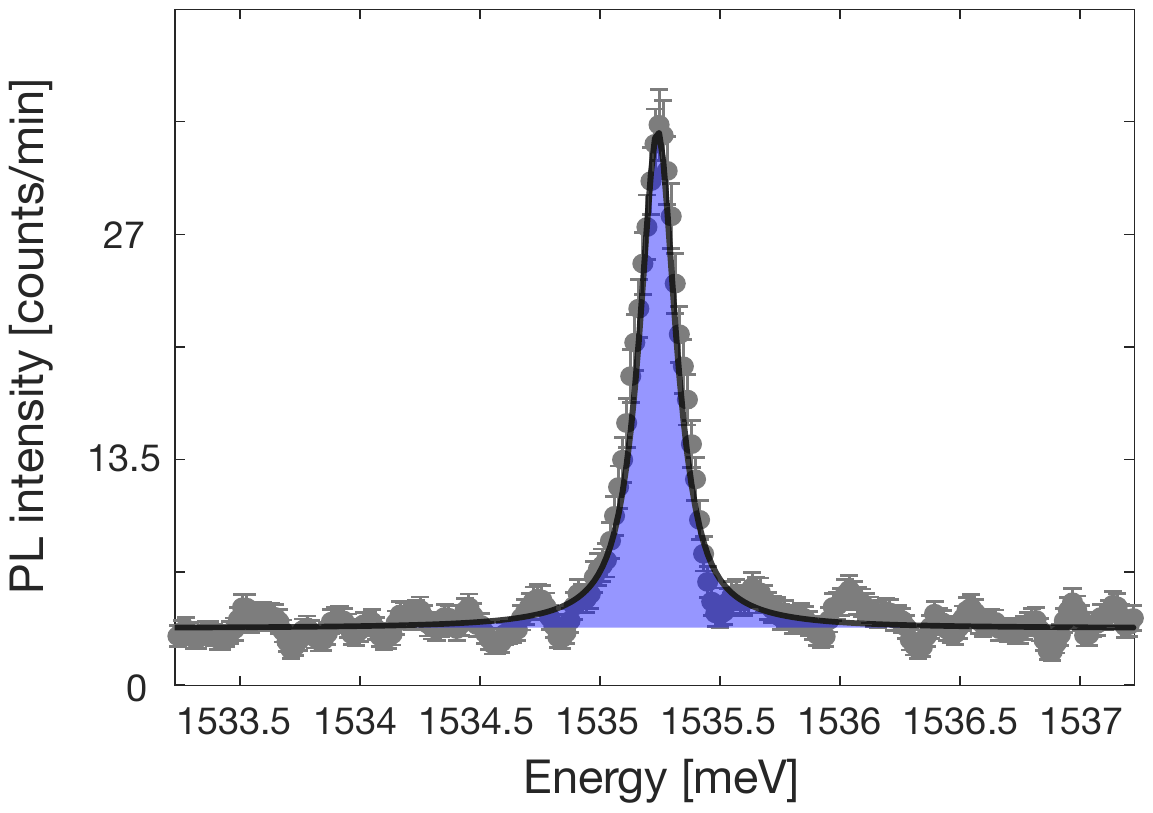}}
\textbf{\fd{Extended data Fig.2}}: \fd{\textbf{Evaluation of the residual doping level.}} 
\fd{PL spectrum radiated by neutral dipolar excitons for $\nu_X\approx1/2$ (at $\nu_{CX}\approx0$). The spectrum is measured by averaging 10 realisations performed under unchanged conditions. The profile is given by our spectral resolution, i.e. reproduced by a single lorentzian-like line with around 150 $\mu$eV full-width-at-half-maximum (blue area and black line). Measurements were performed at 330 mK, error bars displaying the level of poissonian fluctuations.}

\newpage

\centerline{\includegraphics[width=.9\linewidth]{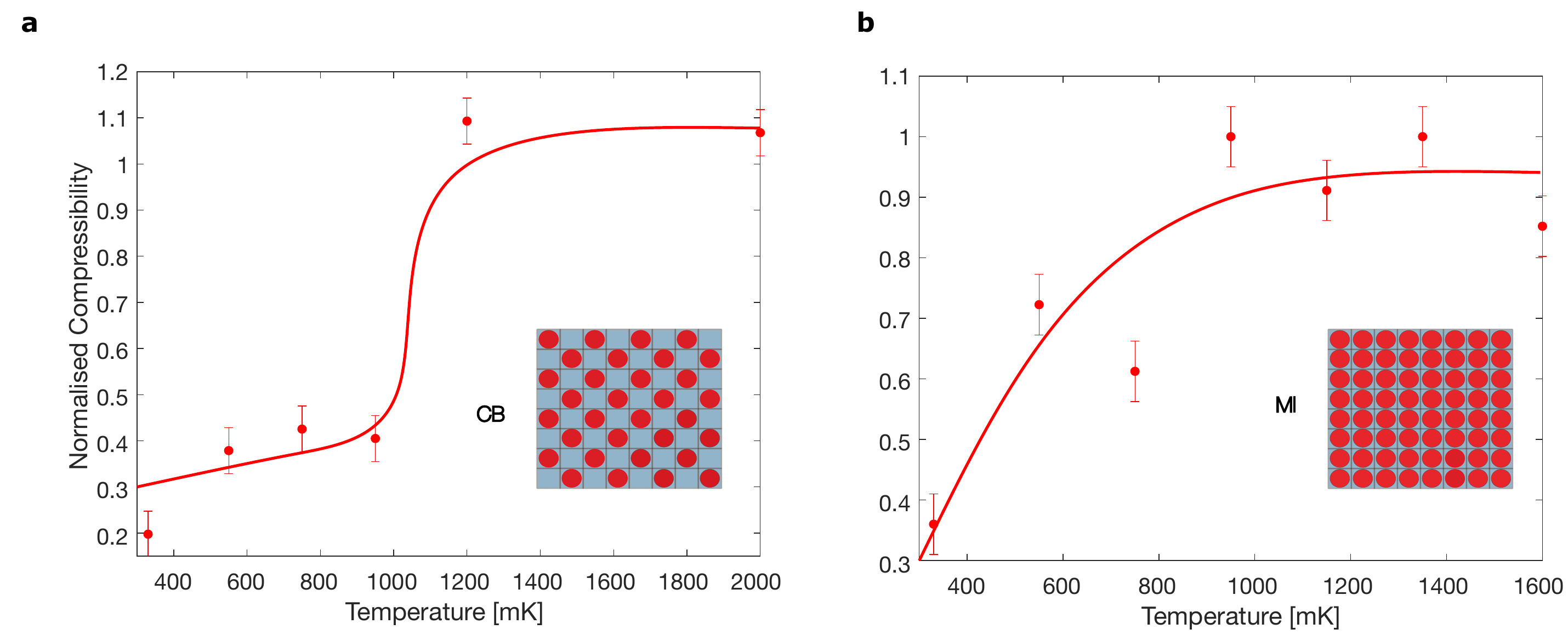}}
\textbf{\fd{Extended data Fig.3}}: \fd{\textbf{Thermal melting of CX insulators at $\nu_{CX}=1/2$ and 1.}} 
\fd{ \textbf{a} Compressibility $\kappa_{CX}$ normalised to the level given by poissonian noise for ($\nu_{CX}=1/2$, $\nu_X\approx0$) as a function of the bath temperature. \textbf{b} Identical measurements for ($\nu_{CX}=1$, $\nu_X\approx0$) . While in \textbf{a} the thermal melting of the insulating phase occurs around 1K, as expected for the magnitude measured for $V_{CX,CX}$, a similar critical temperature is found in \textbf{b} for the Mott phase. This possibly reflects fluctuations of the density of injected holes while the bath temperature is increased. For all measurements error bars mark our statistical precision when computing the compressibility ($\pm0.03$).}

\newpage

\centerline{\includegraphics[width=\linewidth]{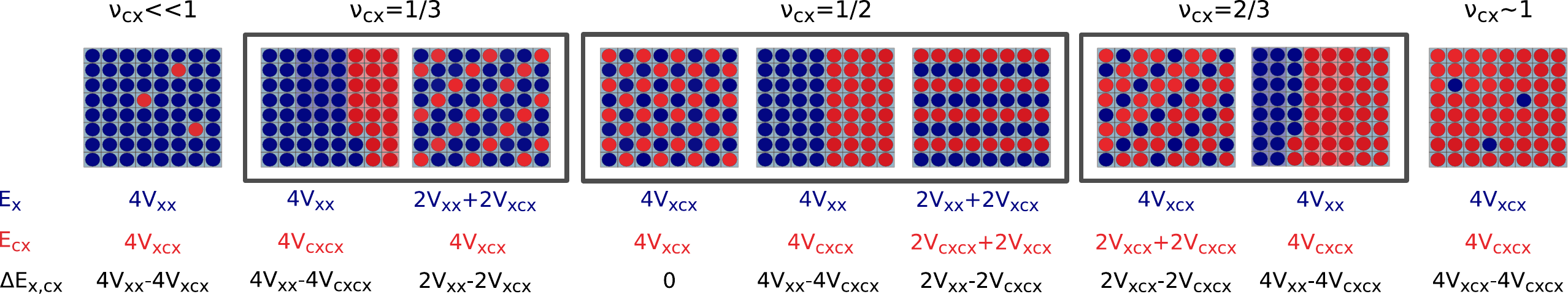}}
\textbf{\fd{Extended data Fig.4}}: \fd{\textbf{Interaction energies and spatial ordering.}} 
\fd{Possible configurations of incompressible phases made by neutral (blue) and charged (red) excitons. The respective energy shifts of PL energies, $E_{X}$ and $E_{CX}$, are indicated below each configuration together with the resulting magnitude of $\Delta E_{X,CX}$, by  only  taking into account NN interactions.}

\end{document}